\documentclass[
 reprint, superscriptaddress,
 amsmath,amssymb,
 aps,longbibliography,
prb,
floatfix
]{revtex4-2}
 
\usepackage{dcolumn}
\usepackage{bm}

\usepackage{hyperref}
\hypersetup{
    colorlinks=true,
    linkcolor = [rgb]{0.70,0.13,0.13}, 
    citecolor = [rgb]{0.13,0.55,0.13},
    urlcolor = [rgb]{0.25, 0.41, 0.88} , 
    filecolor=cyan,      
    pdfcreator = {\LaTeX\ and \flqq hyperref\frqq},
}

\usepackage{cleveref}

\RequirePackage{graphicx,amsmath,amssymb,bm}
\usepackage{graphicx}
\usepackage{dcolumn}
\usepackage{bm}
\usepackage{appendix}
\RequirePackage{graphicx,amsmath,amssymb,bm}
\usepackage{bbm}
\usepackage{framed} 
\usepackage{amsthm}
\usepackage{caption}
\usepackage{subcaption}
\usepackage[normalem]{ulem}
\usepackage{comment}
\usepackage{latexsym}
\usepackage{amssymb}
\usepackage{amsmath}
\usepackage{amsfonts}
\usepackage[vcentermath]{youngtab}
\usepackage{upgreek}
\usepackage{bm,dsfont}
\usepackage{multirow}
\usepackage{braket}
\usepackage{enumitem}
\usepackage[percent]{overpic}
\usepackage[usenames, dvipsnames]{color}
\usepackage[svgnames]{xcolor}

\newcommand{\up}{\uparrow}
\newcommand{\dn}{\downarrow}

\graphicspath{ {./Figs/} }

\newcommand{\Tr}{\text{Tr}}

 \newcommand{\e}{\text{e}}

\newcommand{\bk}{\bm{k}}

\def\be{\begin{equation}}
\def\ee{\end{equation}} 
\def\bsh{\begin{shaded}}
\def\esh{\end{shaded}} 
\def\bpm{\begin{pmatrix}}
\def\epm{\end{pmatrix}}

\begin{document} 
\preprint{APS/123-QED}
\title{Entanglement spectrum of gapless topological phases: a case study with topological superconductors} 
\author{Archi Banerjee}
\affiliation{Max Planck Institute for the Physics of Complex Systems, Nöthnitzer Str. 38, 01187 Dresden, Germany}
\affiliation{Max Planck Institute for Chemical Physics of Solids, Nöthnitzer Str. 40, 01187 Dresden, Germany}
\affiliation{SUPA, School of Physics and Astronomy, University of St Andrews, St Andrews KY16 9SS, United Kingdom}
\author{Meng Zeng} 
\email{mengzeng@pks.mpg.de}
\affiliation{Max Planck Institute for the Physics of Complex Systems, Nöthnitzer Str. 38, 01187 Dresden, Germany}

\begin{abstract} 
Using bulk gapless topological superconductors in both 1d and 2d as free fermion model examples, we demonstrate the power of subsystem correlation spectrum (the spectrum of correlation matrix), or equivalently the entanglement spectrum for the case of free fermions, in characterizing the topology of the non-trivial ground state. For the systems considered, we show that signatures of the low-energy spectrum, including both the edge modes and the bulk modes, appear in the correlation spectrum, albeit with different behaviors. This work generalizes the 2d Li-Haldane entanglement spectrum characterization of topological edge states to 2d topological systems with gapless bulk. 
\end{abstract} 
\maketitle

\section{Introduction} 
Quantum entanglement is a powerful tool in condensed matter physics for probing quantum correlations and characterizing phases of matter \cite{vedral-rmp-2008}. It offers key insights at quantum critical points and detects topological order through measures like topological entanglement entropy (EE) \cite{levin-wen-2006,kitaev-preskill-2006}. 
After the pioneering work of Li and Haldane \cite{li_haldane}, where the authors identified topological signatures of  fractional quantum Hall states in the entanglement spectrum (ES), it has been realized that the ES contains more physical information than just the EE alone. 

This insight was soon generalized to symmetry protected topological (SPT) phases \cite{pollmann_ES,turner_inversion}, where the degeneracy in the bulk ES is shown to be a more robust characterization of the non-trivial topology than the existence of edge modes. For generic cases with stable topological edge modes, the correspondence between the low energy modes of the bulk ES and the edge mode spectrum have been proved in various settings, including free fermion systems \cite{fidkowski_entanglement_2010}, 
fractional quantum Hall systems \cite{anushya-proof-FQH}, Kitaev model \cite{yao-qi-2010}, and generic interacting systems \cite{Peschel-chung-2011,qi_proof}. Recent advancements in cold atom experiments have made it possible to directly measure the ES of many-body systems \cite{pichler2016measurement,dalmonte2018quantum,Dalmonte_2022}.
While the power of ES in characterizing various physical properties has been demonstrated abundantly in the literature, it has also been pointed out \cite{anushya-2014} that the ES can contain non-universal information for symmetry breaking phases \cite{masud-ssb,ssb-2,ssb-3} and topological phases with edge reconstruction \cite{edge-reconstruction}, hence has to be handled with caution. 

In recent couple of years, there have been intensive studies in topological phases with a gapless bulk, i.e. there are robust symmetry protected edge modes even when the bulk is gapless \cite{keselman,verresen-jones,scaffidi_gapless_2017,verresen_gapless_2021}. A natural question to ask is whether the diagnostic of the topology through ES degeneracy still works for gapless topological phases \cite{calabrese_entanglement_2008,pollmann_ES_critical,thomale-momentum-space,thomale-nonlocal-order,thomale-universal-ES}. Some recent works have pursued this question and have demonstrated the effectiveness of ES degeneracy in identifying non-trivial topology in certain 1d critical systems \cite{yu_ES,zhong2025quantum}. Ref.~\cite{scaffidi_gapless_2017} also provided a 2d construction of interacting gapless SPT, where the authors showed signs of an extra topological edge mode in the gapless non-trivial phase using exact diagonalization. In this work, we aim to provide more insight to the challenging 2d case, using an exactly solvable free-fermion system, the nodal $d$-wave topological superconductor (TSC) \cite{TSC-2010,majorana-flat-2013}, whose entanglement properties can be calculated exactly using two-point correlation functions \cite{cheong_many-body_2004,peschel_calculation_2003,peschel_reduced_2009,Peschel-chung-2011}. More specifically, we calculate the correlation spectrum (CS), sometimes also termed single-particle ES \cite{trace-index}, which is defined as the spectrum of the subsystem two-point correlation matrix. Based on the CS, the many-body ES can be readily calculated. It has been shown numerically that in the bulk gapped case, spectral flow in the CS signifies the non-trivial topology of the system \cite{trace-index}. 
In a proof for gapped free fermion systems by Fidkowski, a crucial ingredient is spectral flattening \cite{fidkowski_entanglement_2010}, where all the gapped bulk energy bands are flattened such that the full bulk energy spectrum collapses to $\pm 1$ based on the sign of the energy eigenvalues. As a result, when there is an open boundary, only the edge states can appear as low energy modes in the spectrally flattened Hamiltonian. 

In contrast, for systems with gapless bulk, the procedure of spectral flattening is no longer well-defined in the thermodynamic limit. Therefore, it is not obvious what information one can extract from the CS in this case. 
In the 2d nodal TSC with $d_{x^2-y^2}$ pairing that we considered, the CS also shows spectral flow, indicating the existence of chiral Majorana zero modes on the edge. For the $d_{xy}$ pairing case without edge Majorana zero modes, but still topologically non-trivial, the spectral flow is cut off due to the existence of bulk nodal point with the same momentum at the Fermi level. Nevertheless, in both cases we can see clear signatures of low-energy modes from both the edge and the bulk, but with quite different behaviors. We also calculate the trace index, defined as the average subsystem particle number \cite{trace-index}, as a complementary quantification of the spectral flow in the CS.

The paper is organized as the following: in Sec.~\ref{sec:ES-analytics} we review the basics of the ES and its derivation from the CS in free fermion systems. In Sec.~\ref{sec:1d}
we reproduce some known results in 1d topological critical Majorana chains for completeness and further comment on the inapplicability of the spectral flattening proof by Fidkowski in gapless systems by providing some scaling arguments. Then we provide detailed calculations for 2d $d$-wave nodal TSC, including the CS and the trace index, in Sec.~\ref{sec:2d}. Finally we conclude in Sec.~\ref{sec:conclusion} and mention some open questions for future study.
\section{CS and ES for free fermion systems}
\label{sec:ES-analytics}
Given any pure state $\ket{\Psi}$ defined on the total space $A\cup \bar{A}$, where $A$ is the chosen subsystem and the $\bar{A}$ is the complement, the reduced density matrix on $A$ is defined as $\rho_A\equiv\Tr_{\bar{A}}\rho$, with $\rho\equiv\ket{\Psi}\bra{\Psi}$ being the total density matrix of the system. The reduced density matrix $\rho_A$ encodes the entanglement information between $A$ and $\bar{A}$. In particular, the von Neumann EE between the two is given by $S_A=-\Tr\rho_A\ln \rho_A$. However, more detailed information regarding entanglement can be obtained by looking at the ES, defined as the spectrum of the so-called entanglement Hamiltonian $H_E\equiv -\ln \rho_A$ \cite{li_haldane}.

Generally speaking, the calculation of the ES for a many-body ground state of an interacting Hamiltonian is a non-trivial task, at least analytically. However, for free fermion systems, an exact analytical treatment is possible \cite{chung_density-matrix_2001,peschel_calculation_2003,peschel_reduced_2009,fidkowski_entanglement_2010,cheong_many-body_2004}. In this section we review briefly the analytical calculation for the ES and the EE for free fermion systems, possibly with pairing terms, using two-point correlation functions. The basic idea is that, for free fermion systems all the physical information is contained in two-point correlations because higher-point correlations can always be decomposed into two-point correlations by Wick's theorem. On the other hand, the reduced density matrix $\rho_A$, which can be used to calculate all the physical observables in the subsystem $A$, also contains information about all the two-point correlations. Therefore, the idea is to reconstruct the $\rho_A$ from the fermion two-point correlation functions. Below we sketch out the basic workflow. 

Consider now the following generic free-fermion Hamiltonian consisting of all the fermion bilinear terms, including pairing:
\begin{equation}\label{eqn:Ham1}
    H = \sum_{ij} \left[c^\dagger_i A_{ij} c_j + \frac{1}{2} \left(c^\dagger_i B_{ij} c^\dagger_j + \text{h.c.}\right)\right],
\end{equation}
where $c_i$ is the fermion annihilation operator with the collective index $i$ containing information about site, spin and orbital etc. Hermiticity of $H$ requires $A_{ij}=A_{ji}^*$ and $B_{ij}=-B_{ji}$. Eq.~(\ref{eqn:Ham1}) can be straightforwardly diagonalized by defining the Nambu basis $\psi=(c_1,c_1^\dagger,c_2,c_2^\dagger,..,c_N,c^\dagger_N)^T$, where $N$ is the total number of degrees of freedom. After diagonalization the quasiparticle operator will be given by a combination of particle and hole operators $\eta_k = \sum_i u_{ki} c_i + v_{ki} c^\dagger_i$ such that 
$ H = \sum_k \epsilon_k \eta^\dagger_k \eta_k + \text{const.}$

Owing to the particle hole-symmetry in the BdG system, the energy spectrum is symmetric about zero energy. Therefore, the ground state of this Hamiltonian is constructed by filling the negative energy states \cite{noauthor_superconductivity_2007,chung_density-matrix_2001}:
\begin{equation}
    \ket{\Psi} = \prod_{\epsilon_k<0}\eta^\dagger_k\ket{0},
\end{equation}
where $\ket{0}$ is the vacuum of the hole operator $c_i\ket{0}=0$. Given the ground state, the two-point correlation functions $C_{ij}=\braket{c^\dagger_ic_j}$ and $F_{ij}=\braket{c^\dagger_ic^\dagger_j}$ can now be readily calculated. Based on the two types of correlations, the full correlation matrix $G$ is given by
\begin{equation}
    G_{ij} = \begin{pmatrix}
        \braket{c^\dagger_ic_j} & \braket{c^\dagger_ic^\dagger_j}\\
        \braket{c_ic_j} & \braket{c_ic^\dagger_j}
    \end{pmatrix} 
    = \begin{pmatrix}
        C_{ij} & F_{ij} \\
        F^\dagger_{ij} & \delta_{ij}-C_{ji}
    \end{pmatrix}.
\end{equation}
It has been shown that the reduced density matrix for the subsystem $A$ can be inferred from the $G_A$ \cite{peschel_reduced_2009,zhong2025quantum}, i.e. the correlation matrix restricted to $A$. More precisely, 
\begin{equation}
    \rho_A =  \det \left(1-G_A\right)\exp\left(\sum_{ij\in A} [\ln G_A(I-G_A)^{-1}]_{ij}\psi^\dagger_i\psi_j\right),
\end{equation}
which gives the direct relation between the entanglement Hamiltonian $H_E=-\ln \rho_A$ and the subsystem correlation matrix $G_A$. The EE can then be expressed in terms of the eigenvalues $\{\xi_n\}$ of $G_A$ as
\begin{equation}
    S_A=-\sum_n\left[\xi_n\ln \xi_n+\left(1-\xi_n\right)\ln \left(1-\xi_n\right)\right].
\end{equation}
Notice that when $\xi_n\to 0,1$, the contribution to the EE vanishes. The maximum contribution comes from $\xi_n\to 1/2$, which is $\ln 2$. Such a $1/2$ mode in the CS is exactly related to the topological edge modes, as proved for bulk gapped topological systems \cite{fidkowski_entanglement_2010}. We aim to show that this also holds in topological systems with a gapless bulk.

\section{CS for 1D topological Majorana chains}
\label{sec:1d}
\subsection{Bulk gapped and bulk critical Majorana chains}
In this section we reproduce some known results for the 1d Majorana chain system \cite{verresen-jones,zhong2025quantum}. 
\begin{figure}[t]
    \centering
   \includegraphics[width=\linewidth]{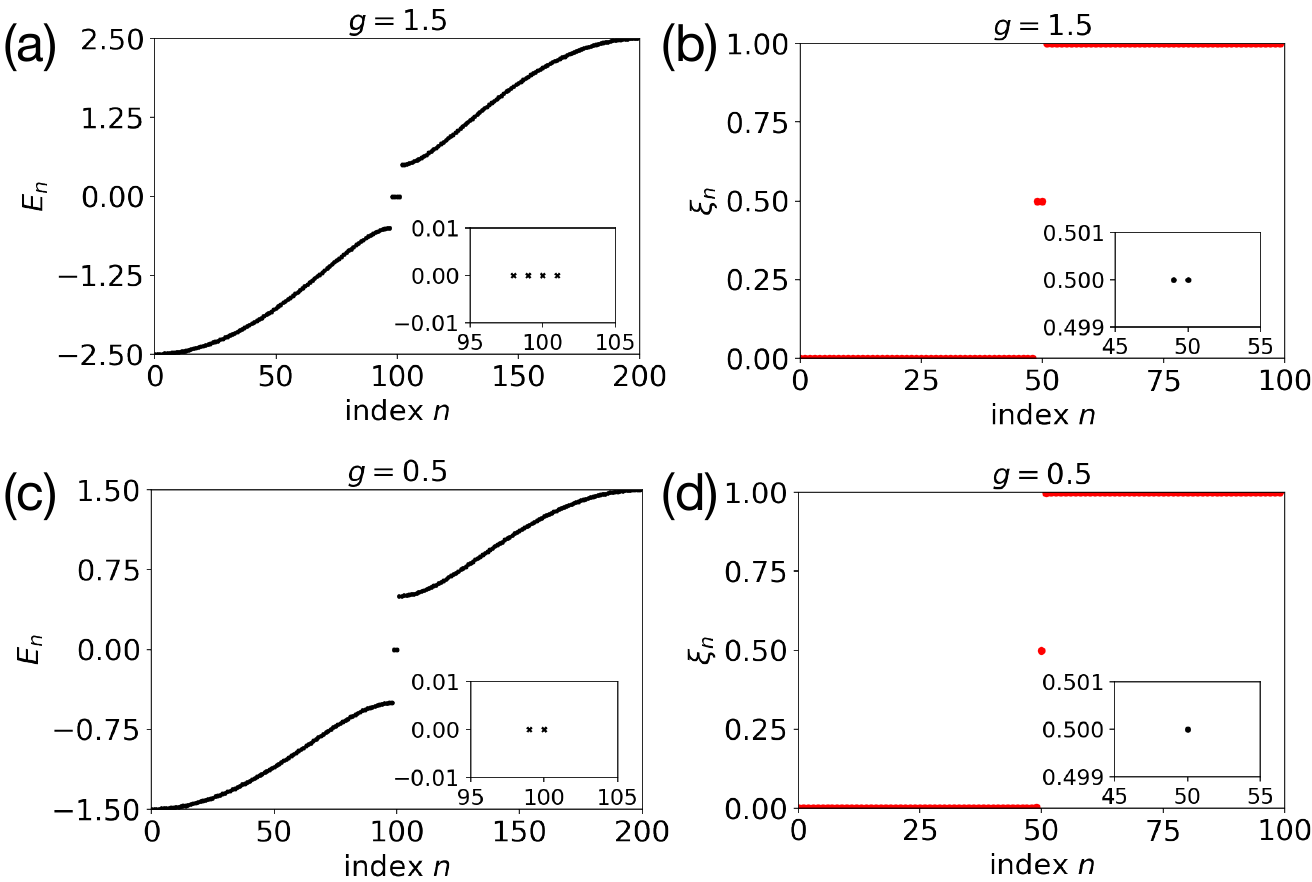}

    \caption{(a) Energy spectrum under open boundary condition (OBC) and (b) correlation spectrum for gapped $\alpha$-chain at $g=0.5$, with winding number $\omega_1=1$; (c) energy spectrum under OBC and (d) correlation spectrum for gapped $\alpha$-chain at $g=1.5$, with winding number $\omega_2=2$. The length of the chain is taken to be $100$ unit cells and correlation spectra are calculated for a single virtual cut on the open chain dividing it into two subsystems.}
    \label{fig:gapped_1D}
\end{figure}
\begin{figure}[t]
    \centering
    \includegraphics[width=\linewidth]{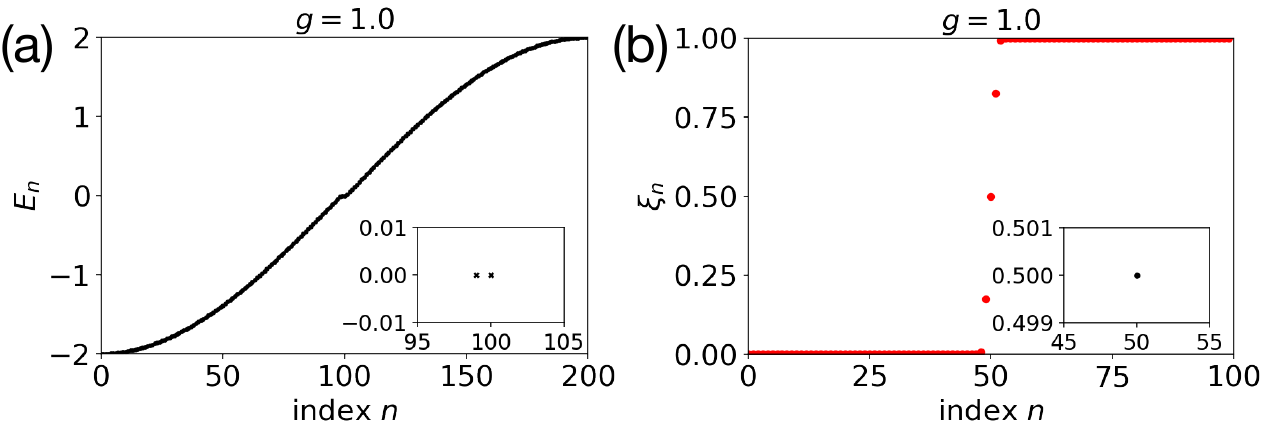}

    \caption{(a) Energy spectrum and (b) half-chain correlation spectrum under OBC for the critical point corresponding to $g=1$. }
    \label{fig:critical1D}
\end{figure}
We consider the Hamiltonian $H=H_1 - gH_2$, where $H_\alpha$ is the so-called $\alpha$-chain Hamiltonian \cite{verresen-jones}:
\begin{equation}
    H_\alpha = i \sum_n \tilde{\gamma}_n \gamma_{n+\alpha}
\end{equation}
The $\alpha$-chain preserves particle hole symmetry $\mathcal{C}$ and time reversal symmetry $\mathcal{T}
$ with $\mathcal{T}^2=\mathcal{C}^2=1$, and thus belongs to the BDI symmetry class. If we write out the Hamiltonian in terms of complex fermion operators using $\gamma_n=c^\dagger_n+c_n$ and $\tilde{\gamma}_n=i(c^\dagger_n-c_n)$, we obtain the Hamiltonian in the complex fermion basis:
\begin{equation}
\begin{aligned}
    H = -\sum_n \left[(c^\dagger_nc^\dagger_{n+1} + c^\dagger_n c_{n+1} + h.c.)
    \right.\\ 
\left.  
    -g(c^\dagger_nc^\dagger_{n+2} + c^\dagger_n c_{n+2} + h.c.)\right],
\end{aligned}
\end{equation}
which is a free complex fermion model with pairing terms. It is known that there are two quantum phase transitions at $g=\pm 1$ \cite{verresen-jones}. The winding number for $|g|<1$ is $1$ and for $|g|>1$ is $2$.  For this bulk gapped case away from the critical points, we find $1$ or $2$ Majorana zero modes on each end of an open chain, resulting in double or quadruple degeneracy in the energy spectrum with open boundary condition for $g=0.5,1.5$ (see Fig.~\ref{fig:gapped_1D}(a)(c)). 
The corresponding CS for the two cases are as shown in Fig.~\ref{fig:gapped_1D}(b)(c), where the number of in gap $1/2$ modes for one virtual cut agrees with the number of edge Majorana zero modes.
This agreement of the topological winding number, degeneracy of OBC energy spectrum and degeneracy in the CS is expected for the bulk gapped phase, as shown by Fidkowski \cite{fidkowski_entanglement_2010}.

Now we move on to the bulk gapless critical point. According to \cite{verresen-jones},  two gapped $1+1$D topologically non-trivial phases of the $\alpha$-chain, with winding numbers $\omega_1>\omega_2>0$ are separated by a critical point with $\omega_2$ topologically protected edge modes on each edge. Thus, we expect the critical $\alpha$-chain $H$ with open boundary conditions at $g=\pm1$ to always have a two-fold degeneracy in the energy spectrum under OBC and one $1/2$-mode in the subsystem CS with one virtual cut. Indeed, we confirm this expected behavior in Fig.~\ref{fig:critical1D} for $g=1.0$.

\subsection{Scaling of CS degeneracy splitting and comment on spectral flattening}
\begin{figure}[t]
    \centering
    \includegraphics[width=\linewidth]{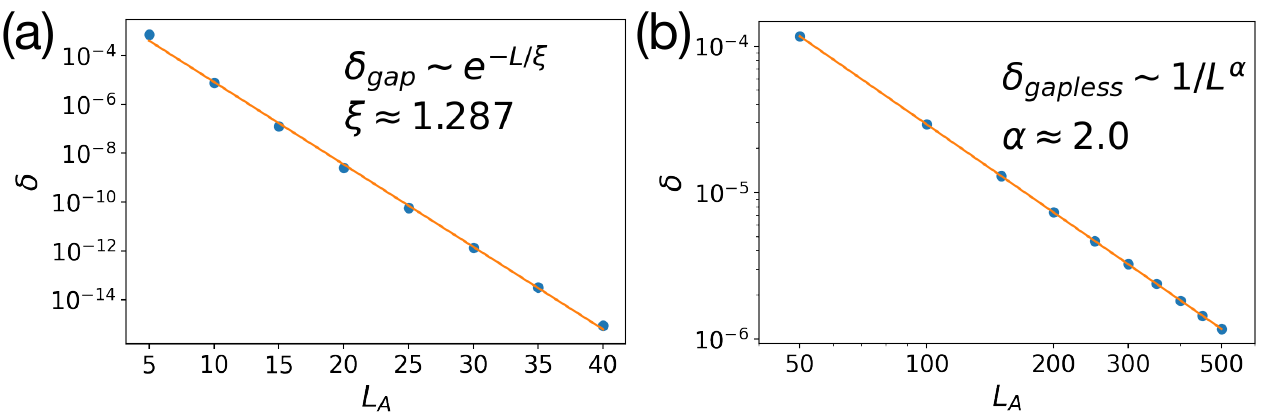}
    \caption{Scaling of degeneracy splitting $\delta$ between the $1/2$ modes of the correlation spectrum with respect to subsystem size $L_{A}$ for \textit{two} cuts on an open chain for (a) the gapped $\alpha$-chain with $g=0.5$ and (b) the critical $\alpha$-chain with $g=1.0$.}
    \label{fig:edgeScaling}
\end{figure}

\begin{figure*}[!t]
    \centering
    \includegraphics[width=0.9\textwidth]{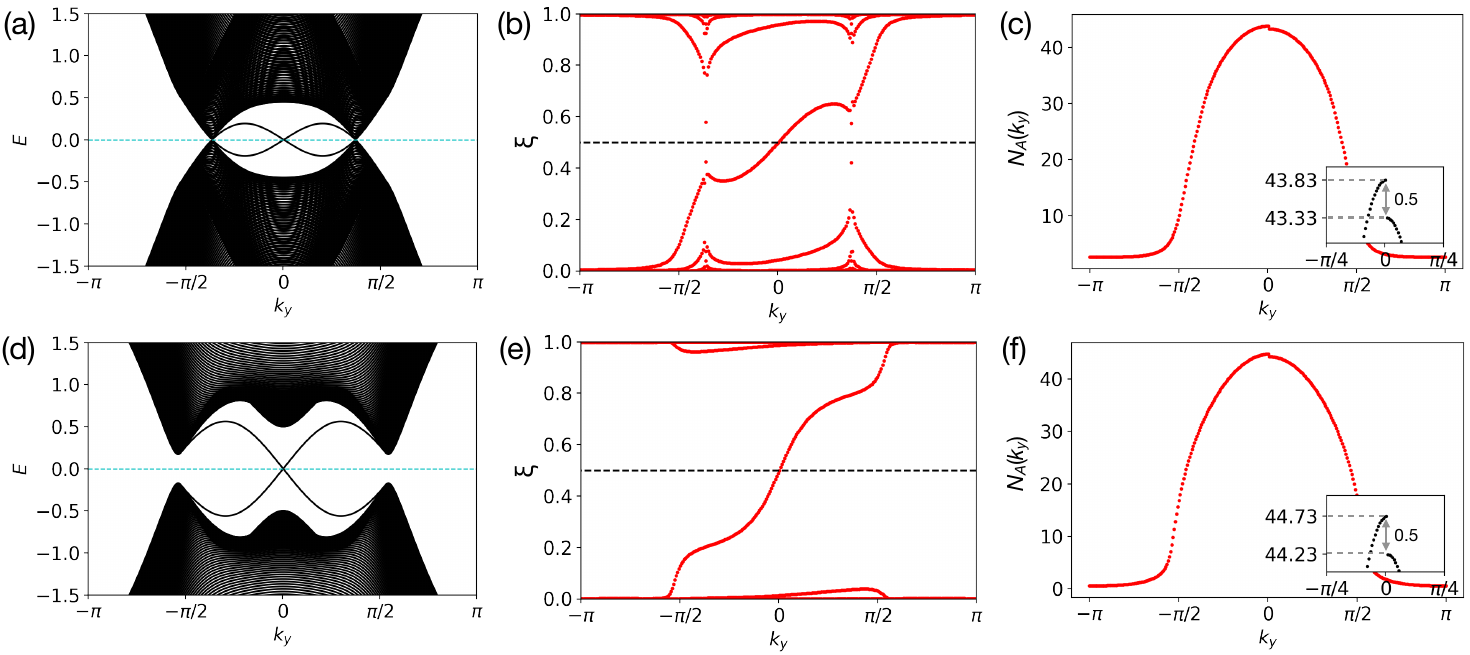}
       \caption{(a) Energy spectrum, (b) CS and (c) trace index of the nodal $d_{x^2-y^2}$-wave TSC. (d)(e)(f) show the case with gapped nodes respectively. $L_x=160$, $L_y=300$, PBC along $y$-direction and OBC along $x$-direction with one virtual cut on the cylinder for correlation spectrum and trace index between unit cells 76 and 77. Parameters: $t=1$, $\mu=-4t$, $\lambda=0.5t$, $h=2t$, $\Delta=t$ for bulk gapless case. To gap out bulk gapless nodes, we use a $C_4$-breaking term $d(\tau_y\otimes\sigma_y)$, with $d=1.5$ and $\tau_y$ being the Pauli matrix in particle-hole space, reducing it to class $D$ of the ten-fold way which has a Chern number classification.}
    \label{fig:dx-y}
\end{figure*}

In this section, we look deeper into the finite-size scaling of the degeneracy splitting of the subsystem correlation spectrum (CS), to gain more insight into the relation between the subsystem correlation matrix and the subsystem Hamiltonian in the bulk gapless case. For this purpose, we take two cuts on the open $\alpha$-chain placed symmetrically from the two ends of the open chain. This yields a two fold degeneracy in the correlation spectrum, arising from a single topological edge mode from each edge.

As a comparison, we firstly show the CS degeneracy splitting for the bulk gapped $\alpha$-chain with $g=0.5$ in Fig.~\ref{fig:edgeScaling}(a). The splitting decays exponentially, similar to the behavior of energy splitting of degenerate physical edge modes. This is consistent with Fidkowski's proof \cite{fidkowski_entanglement_2010}, where it was shown that the subsystem correlation matrix is the same with the subsystem Hamiltonian of a spectrally flattened bulk.

In contrast, for the topological critical chain at $g=1$ shown in Fig.~\ref{fig:edgeScaling}(b), we see that the degeneracy splitting of the subsystem CS follow a power law with the subsystem size $L_A$, i.e. $\sim 1/L_A^\alpha$, which is different from the behavior of the energy splitting of the physical edge modes at the system boundary. Therefore, for the critical chain, the subsystem correlation matrix $G_A$ (see Sec.~\ref{sec:ES-analytics}), and consequently the corresponding entanglement Hamiltonian $H_E$, is a fundamentally different object from the physical subsystem Hamiltonian $H_A$.  This provides direct numerical evidence for the inapplicability of Fidkowski's spectral flattening argument in the bulk gapless case.

However, it has to be pointed out that, in the context of conformal field theory it is known that the natural length scale is $\log L_A$. With this length scale, the degeneracy splitting of the CS decays exponentially, i.e. $\sim \e^{-\alpha \log L_A}$, and the corresponding bulk gap scaling in CS follows a power law in $\log L_A$ \cite{yu_ES,yu2025generalized}.

\section{2D nodal topological superconductors}
\label{sec:2d}
As a 2D model system, we consider the nodal $d$-wave topological superconductor with $4$-by-$4$ BdG Hamiltonian in the basis $\psi^\dagger=(c_{\bk,\up}^\dagger,c_{\bk,\dn}^\dagger,c_{-\bk,\up},c_{-\bk,\dn})$ \cite{majorana2010}:
\begin{equation*}
    H(\bk) = \begin{pmatrix}
        \epsilon_{\boldsymbol{k}}- h\sigma_z + \boldsymbol{g}_{\boldsymbol{k}}\cdot\boldsymbol{\sigma} & i \Delta_{\boldsymbol{k}} \sigma_y  \\
        - i\Delta_{\boldsymbol{k}} \sigma_y & -\epsilon_{\boldsymbol{k}} + h\sigma_z + \boldsymbol{g}_{\boldsymbol{k}} \cdot \boldsymbol{\sigma}^*
    \end{pmatrix},
\end{equation*}
where $\epsilon_{\bk}$ is the electron dispersion, $h$ is the Zeeman field, $\Delta_{\bk}$ is the pairing function and $\boldsymbol{\sigma}=(\sigma_x,\sigma_y,\sigma_z)$ represent the Pauli matrices for the spin-$1/2$ degree of freedom. 
Here, $\epsilon_{\bk} = -2t(\cos k_x+\cos k_y)-\mu$ and $\boldsymbol{g}_{\bk} = 2\lambda(\sin k_y,-\sin k_x,0)$, with $t$ being the hopping amplitude, $\mu$ being the chemical potential and $\lambda$ characterizing the spin-orbit coupling. 
We will be considering two types of $d$-wave pairings: $\Delta_{\bk} = \Delta (\cos k_x-\cos k_y)$ for $d_{x^2-y^2}$ pairing and $\Delta_{\bk} = \Delta \sin k_x \sin k_y$ for $d_{xy}$ pairing, with $\Delta$ being the pairing amplitude. The system is in a topologically non-trivial phase for $-4t-\mu<h<-\mu$.  
We consider a cylindrical geometry with periodic boundary conditions along the $y$-direction and open boundary conditions or virtual cuts along the $x$-direction for calculating our results. Below we show the results for the two different pairings separately, but it has to be pointed out that both the (low) energy spectra and the (low) correlation spectra of the two pairings are related to each other by a $45^o$ rotation. This can be straightforwardly checked by repeating the calculation on a rotated lattice, while keeping the cut direction the same as before.

\subsection{\texorpdfstring{$d_{x^2 - y^2}$ pairing}{dx2-y2 pairing}}

\begin{figure*}[!t]
    \centering
     \includegraphics[width=0.9\textwidth]{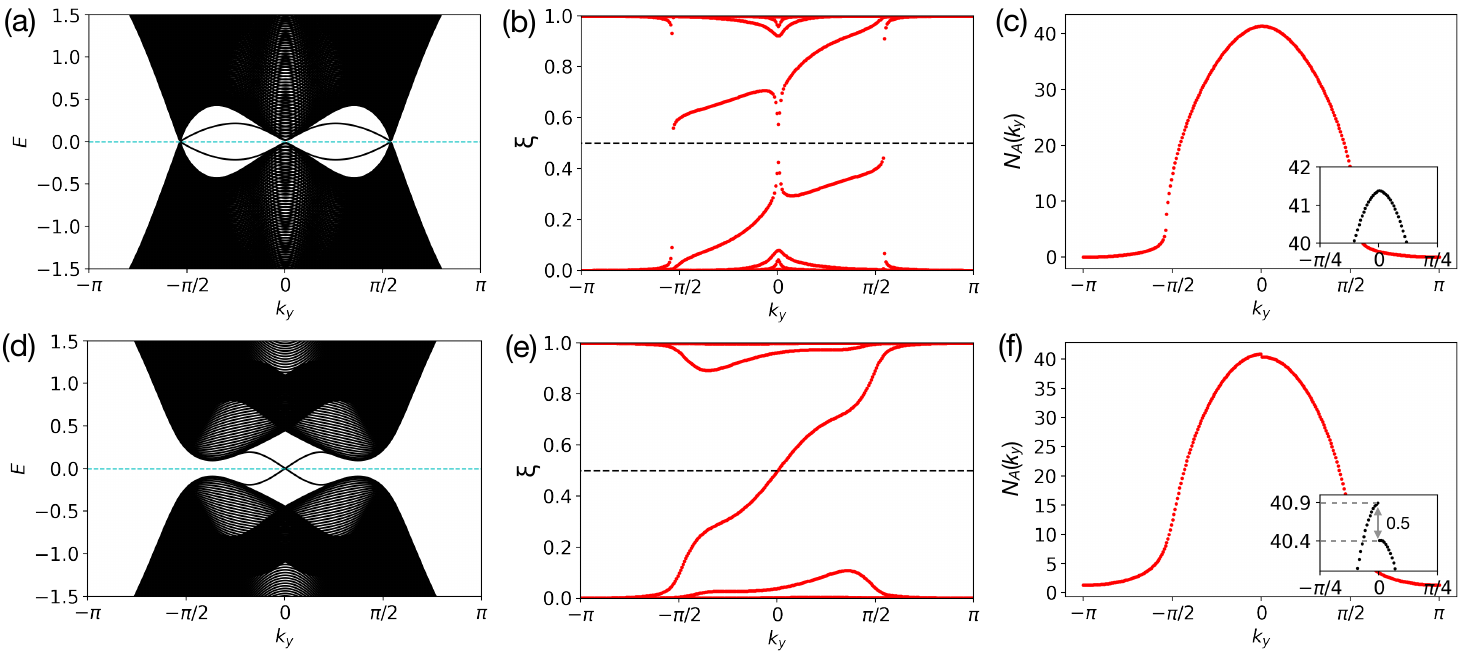}
       \caption{(a) Energy spectrum, (b) CS and (c) trace index of the nodal $d_{xy}$-wave TSC. (d)(e)(f) show the case with gapped nodes respectively. $L_x=160$, $L_y=300$, PBC along $y$-direction and OBC along $x$-direction with one virtual cut on the cylinder for correlation spectrum and trace index between unit cells 76 and 77. Parameters: $t=1$, $\mu=-4t$, $\lambda=0.5t$, $h=2t$, $\Delta=t$ for bulk gapless case. To gap out bulk gapless nodes, we use a $C_4$-breaking term $d(\tau_y\otimes\sigma_y)$, with $d=1$ and $\tau_y$ being the Pauli matrix in particle-hole space, reducing it to class $D$ of the ten-fold way which has a Chern number classification.}
    \label{fig:dxy}
\end{figure*}
In the topologically non-trivial phase of the model with $d_{x^2-y^2}$ pairing, the chiral Majorana edge modes appear on opposite edges of the cylinder, where both modes cross the Fermi level at $k_y=0$ (see Fig.~\ref{fig:dx-y}(a)). The bulk nodes show up in the energy spectrum near $k_y=\pi/2$, where the edge modes merge into the gapless bulk. Due to the mismatch between the gapless bulk momentum and the gapless edge momentum, the edge modes do not couple to the bulk gapless states in the absence of external momentum transfer mechanism, e.g. existence of impurities, and the Majorana edge modes remain stable \cite{majorana2010}. 
We show in Fig.~\ref{fig:dx-y}(b) evidence of a chiral spectral flow in the neighborhood of $k_y=0$ in the CS of a subsystem with a single cut on the cylinder. At $k_y=0$, the correlation spectrum yields a single mode sitting at $1/2$, corresponding to the Majorana zero mode at the virtual cut. 

Other than the distinctive chiral edge mode shown in the CS, one can also see features of the the bulk gapless nodes around $k_y\sim \pi/2$ in Fig.~\ref{fig:dx-y}(b). Most of the higher-energy bulk modes in the energy spectrum show up as the modes corresponding to the eigenvalues $\sim 0$ or 1 in the CS, which have little contribution to the subsystem EE. However, the low-energy bulk modes, due to their extended nature such that they can strand over the virtual cut and entangle the degrees of freedom on both sides even deep into the bulk, can contribute significantly to the subsystem EE. This can be clearly seen in Fig.~\ref{fig:dx-y}(b), where the bulk bands near the nodes deviate from 0 and 1, and approach the value of $1/2$. It has to be mentioned that the finite gap in the CS around the bulk gapless point is in fact a finite-size effect. The gap decays to 0 in the thermodynamic limit, following a power law of the logarithm of the subsystem size \cite{yu_ES}.

As a complementary indicator of the Majorana zero mode, we also calculate the trace index invariant in our case, following the discussions in \cite{trace-index} on Chern insulators. The trace index is defined as the total fermion number of a subsystem $A$ in a given momentum sector - $N_A(k_y)= \langle \sum_{i\in A} c^{\dagger}_{i,k_y}c_{i,k_y} \rangle$. For $U(1)$ preserving systems, the trace index can be obtained simply by taking the trace of the correlation matrix $C_{ij}=\braket{c^\dagger_i c_j}$. The trace index shows a quantized jump for each chiral edge crossings at the Fermi level. In the case of Chern insulators, each edge mode contributes a jump of one fermion, and the total discontinuity is the Chern number \cite{trace-index}. In the case of BdG systems such as the $2$D TSC, where there is no particle number conservation, the trace index can still be calculated nonetheless. As shown by Fig.~\ref{fig:dx-y}(c), for the topologically non-trivial case, there is a jump of $1/2$ in the trace index at $k_y=0$, indicating the Majorana nature of the chiral edge mode. In contrast, there are no such (half-)quantized discontinuous jumps for bulk gapless points.

For comparison, we also added a $C_4$-breaking term to gap out the bulk nodes, whereas the topological edge modes are still intact. The results are shown in Fig.~\ref{fig:dx-y}(d)(e)(f). Quite naturally, the CS now shows predominantly the edge contribution with the bulk modes cleanly pushed close to 0 or 1.
\subsection{\texorpdfstring{$d_{xy}$ pairing}{dxy pairing}}
For the $d_{xy}$ pairing, the bulk energy gap also closes at $k_y=0$, in contrast to the $d_{x^2-y^2}$ case (see Fig.~\ref{fig:dxy}(a)). As a result, the Majorana edge modes merge into the bulk gapless modes at $k_y=0$, hence there are no more chiral Majorana modes on the edge. Another way of seeing this is that the gap function $\Delta_{\bk}$ vanishes at $k_y=0$, and consequently, the system is in a topologically trivial normal metallic phase in this particular momentum sector. Even without topological edge modes, the system can nevertheless be in a topologically non-trivial phase because a superconducting vortex can still host robust Majorana zero modes when there are odd number of them \cite{majorana2010}.  

Similar to the $d_{x^2-y^2}$ case, we look at the CS of the $d_{xy}$ pairing phase. 
We noticed that the CS in this case in general depends on the location of the subsystem cut. More specifically, the in gap eigenvalues show oscillatory behavior with changing cut positions. In particular, for certain cut positions there will be a pair of degenerate eigenvalues occurring exactly at $1/2$. However, it has to be pointed out that such modes are not related to Majorana edge modes as in the 1d case or the 2d case with $d_{x^2-y^2}$ pairing. A crucial difference is that the 1/2-modes due to true topological edge modes are robust features of the CS and are independent from the location of the cut. Such cut-dependent oscillatory behavior is likely inherited from the oscillatory nature of correlation functions in normal gapless liquid, e.g. those with a Fermi surface.

With the above considerations in mind, we choose a generic cut without such 1/2-modes for demonstration purposes.
Quite similar to what we see in the $d_{x^2-y^2}$ pairing, Fig.~\ref{fig:dxy}(b) shows that the CS contains both low-energy contributions from both the edge and the bulk. The difference is that now there is no more $1/2$-mode at $k_y=0$ due to the absence of Majorana zero mode as discussed earlier. What happens at $k_y\sim 0$ for the $d_{xy}$ pairing is similar to the $d_{x^2-y^2}$ pairing for $k_y\sim \pi/2$, where the edge Majorana becomes merged into the bulk at the tip that is gapped away from $1/2$. Not surprisingly, the trace index shows no half-quantized jump near $k_y=0$. The bulk gapped case with the $C_4$-breaking term presented in Fig.~\ref{fig:dxy}(d)(e)(f) is similar to the $d_{x^2-y^2}$ case.

\section{Conclusions}
\label{sec:conclusion}
In this work we tried to generalize the correspondence between entanglement spectrum (ES) and the topological edge modes in bulk gapped topological phases, originally discovered by Li and Haldane in fractional quantum Hall systems, to topological systems with gapless bulk. For this purpose, we used 2d $d$-wave nodal TSC as a model example and calculated the correlation spectrum (CS), which contains the same information with the ES in the case of free fermions, on the cylinder geometry in various cases. We demonstrated that the CS contains signatures of low-energy modes from both the bulk and edge, with quite different behaviors. Trace index is also calculated, whose half-quantized discontinuity in momentum space indicates exactly the existence of zero edge Majorana modes. As a comparison, we also looked into the corresponding bulk gapped cases by adding in $C_4$-breaking terms to gap out the bulk nodes. 

We noticed in passing that the CS seems to have the ability of ``magnifying'' the difference between the topological edge modes and the bulk modes, even though both appear at the same time in low energy. This feature comes handy when investigating general bulk gapless topological phases.

One interesting question for future study is the effectiveness/robustness of the subsystem CS in characterizing topological phases in the presence of interactions. As we already demonstrated in the free fermion case, the in-gap $1/2$-mode in the CS encodes information about the physical topological edge modes. However, it is not clear if this relation still holds for systems with interactions, even when the interactions are relatively weak. We know that for generic free fermion topological systems, with very few exceptions \cite{int-reduced}, the topological nature can survive at least perturbative interactions due to the robustness of topology. However, the CS by definition only contains information about two-point correlations. It is unclear how important the higher-point correlations are in capturing the topological information when interactions are considered. Another question along this line that one could ask is that for systems whose non-trivial topology comes intrinsically from interactions, i.e. no free-fermion counterpart, then whether the CS characterization fails completely. We leave these questions for future study.

\vspace{0.5cm}
\textit{Note added-} While preparing this manuscript, we noticed a recent work \cite{yu2025generalized} that also studied the entanglement spectrum of higher dimension gapless free fermion systems. The authors studied the topological critical point between topologically inequivalent Chern insulators. Here we focus instead on the non-trivial critical phase of topological nodal superconductors. Our results qualitatively agree with each other.

\section*{Acknowledgment}
We would like to thank Lei Su, Kirill Shtengel, Lun-Hui Hu, Roderich Moessner and Ashley Cook for discussions.

\bibliography{main.bib} 

\end{document}